\documentclass[sigconf]{acmart} %review, anonymous (this option removed) 

\usepackage{subfigure}
\usepackage{subcaption}

\usepackage{amsmath,amsfonts}
\usepackage[algo2e,linesnumbered,ruled,lined,noend]{algorithm2e}
\SetKwInOut{Parameter}{Parameters}
\SetKwInOut{Constraint}{Constraints}
\SetKwFunction{Mining}{Mining}
\SetKwFunction{MiningTwo}{Mining2}
\SetKwFunction{Candidates}{Candidates}
\SetKwProg{Fn}{Function}{}{}
\usepackage{graphicx}
\usepackage{textcomp}
\usepackage{color}
\usepackage{xcolor}
\usepackage{dsfont}
\usepackage{bbm}

\usepackage[dvipsnames]{xcolor}

\usepackage{xspace}
\def\BibTeX{{\rm B\kern-.05em{\sc i\kern-.025em b}\kern-.08em
    T\kern-.1667em\lower.7ex\hbox{E}\kern-.125emX}}
    
\usepackage{multirow} % Required for multirows
\usepackage{booktabs} % For prettier tables
\usepackage{colortbl}

\usepackage[normalem]{ulem}
\useunder{\uline}{\ul}{}

\usepackage{pgf}
\usepackage{enumitem}
\setlist[itemize]{leftmargin=*}
\usepackage{stfloats}

\usepackage{makecell}

\usepackage{bbold}
\usepackage{mathtools}
\usepackage{comment}
\usepackage{microtype}
\usepackage[misc]{ifsym}
\usepackage{url}
\usepackage{xspace}

\SetCommentSty{mycommfont}

\newcommand{\sunwoo}{\textcolor{black}}

\newcommand{\method}{{ItemRAG}\xspace}

\definecolor{sunwoogreen}{RGB}{32, 200, 150}
\definecolor{sunwoogreen2}{RGB}{67, 148, 58}
\definecolor{sunwooyellow}{rgb}{1.0, 1.0, 0.0}
\definecolor{sunwooyellow2}{RGB}{228, 208, 10}

\newcommand{\best}{\cellcolor{sunwoogreen!70}}  %{0.9}
\newcommand{\gframe}[1]{\fcolorbox{white}{sunwoogreen!70}{\strut #1}}

\AtBeginDocument{%
  \providecommand\BibTeX{{%
    \normalfont B\kern-0.5em{\scshape i\kern-0.25em b}\kern-0.8em\TeX}}}

\usepackage{needspace}

\AtBeginDocument{%
  \providecommand\BibTeX{{%
    \normalfont B\kern-0.5em{\scshape i\kern-0.25em b}\kern-0.8em\TeX}}}

\copyrightyear{2026}
\acmYear{2026}
\setcopyright{cc}
\setcctype{by}
\acmConference[SIGIR '26]{Proceedings of the 49th International ACM SIGIR Conference on Research and Development in Information Retrieval}{July 20--24, 2026}{Melbourne, VIC, Australia}
\acmBooktitle{Proceedings of the 49th International ACM SIGIR Conference on Research and Development in Information Retrieval (SIGIR '26), July 20--24, 2026, Melbourne, VIC, Australia}
\acmDOI{10.1145/3805712.3809942}
\acmISBN{979-8-4007-2599-9/2026/07}



\begin{document}

    \title[ItemRAG: Item-Based Retrieval-Augmented Generation for LLM-Based Recommendation
    ]{ItemRAG: Item-Based Retrieval-Augmented Generation \\ for LLM-Based Recommendation}
    
    \settopmatter{authorsperrow=3}
	
    \author{Sunwoo Kim}
	\affiliation{%
    	\institution{KAIST}
            \city{Seoul}
            \country{South Korea}
	}
	\email{kswoo97@kaist.ac.kr}

    \author{Geon Lee}
	\affiliation{%
		\institution{KAIST}
            \city{Seoul}
            \country{South Korea}
	}
	\email{geonlee0325@kaist.ac.kr}
    
    \author{Kyungho Kim}
	\affiliation{%
		\institution{KAIST}
            \city{Seoul}
            \country{South Korea}
	}
	\email{kkyungho@kaist.ac.kr}
    
    \author{Jaemin Yoo}
	\affiliation{%
		\institution{Seoul National University}
            \city{Seoul}
            \country{South Korea}
	}
	\email{jaeminyoo@snu.ac.kr}

	\author{Kijung Shin}
	\affiliation{%
		\institution{KAIST}
            \city{Seoul}
            \country{South Korea}
	}
	\email{kijungs@kaist.ac.kr}
	
    \begin{abstract}

Recently, large language models (LLMs) have been widely used as recommender systems, owing to their reasoning capability and effectiveness in handling cold-start items. 
A common approach prompts an LLM with a target user’s purchase history to recommend items from a candidate set, often enhanced with retrieval-augmented generation (RAG).
Most existing RAG approaches retrieve purchase histories of users similar to the target user; however, these histories often contain noisy or weakly relevant information and provide little or no useful information for candidate items.
To address these limitations, we propose ItemRAG, a novel RAG approach that shifts focus from coarse user-history retrieval to fine-grained item-level retrieval.
ItemRAG augments the description of each item in the target user’s history or the candidate set by retrieving items relevant to each.
To retrieve items not merely semantically similar but informative for recommendation, ItemRAG leverages co-purchase information alongside semantic information. 
Especially, through their careful combination, ItemRAG prioritizes more informative retrievals and also benefits cold-start items.
Through extensive experiments, we demonstrate that ItemRAG consistently outperforms existing RAG approaches under both standard and cold-start item recommendation settings.
Supplementary materials, code, and datasets are provided at \url{https://github.com/kswoo97/ItemRAG}.

% Recently, large language models (LLMs) have been widely used as recommender systems, owing to their strong reasoning capability and their effectiveness in handling cold-start items.
% To better adapt LLMs for recommendation, retrieval-augmented generation (RAG) has been incorporated. 
% Most existing RAG methods are user-based, retrieving purchase patterns of users similar to the target user and providing them to the LLM.
% In this work, we propose \method, an item-based RAG method for LLM-based recommendation that retrieves relevant items (rather than users) from item–item co-purchase histories. 
% \method helps LLMs capture co-purchase patterns among items, which are beneficial for recommendations.
% Especially, our retrieval strategy incorporates semantically similar items to better handle cold-start items and uses co-purchase frequencies to improve the relevance of the retrieved items.
% Through extensive experiments, we demonstrate that \method consistently (1) improves the zero-shot LLM-based recommender by up to 43\% in Hit-Ratio@1 and (2) outperforms user-based RAG baselines under both standard and cold-start item recommendation settings.
    \end{abstract}
	
	\maketitle

    \section{Introduction}
    \label{sec:intro}
    \begin{figure}[t] 
  \centering
  \includegraphics[width=0.75\linewidth]{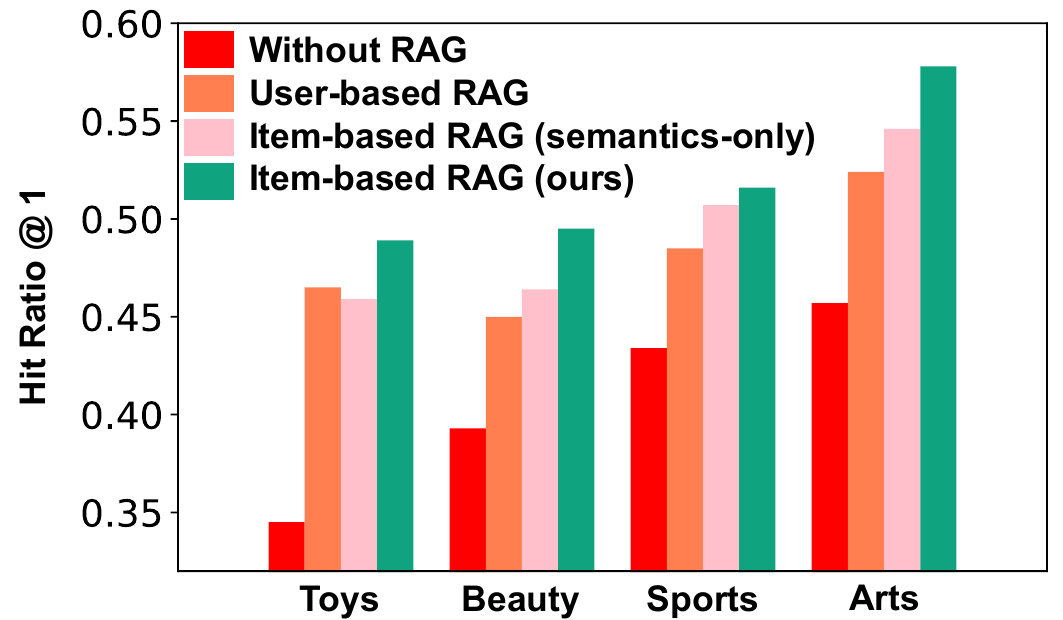} % or .png/.jpg
  \caption{\textit{Strong performance of \method, our proposed item-based RAG method.}
  \method consistently (1) improves the zero-shot LLM-based recommender (Without RAG) and (2) outperforms both the strongest user-based RAG baseline (CoRAL~\cite{wu2024coral}), and a semantics-only item-based RAG method.}
  \label{fig:mainperformance}
\end{figure}

Recommender systems are core to modern web services, retrieving items that match user interests from vast item pools~\cite{kim2026personalized, kim2025multi, gao2025process, liao2025multi, lee2025sealr, kim:hal-05561909, kim2025self, kim2024towards}. 
By inferring users' preferences from their purchase histories, recommender systems provide personalized recommendations that improve user satisfaction and drive business revenue.

In recent years, there have been significant efforts to use \textit{large language models} (LLMs) as a recommender system.
\sunwoo{In standard practice, given a target user’s purchase history, an LLM is prompted to recommend items from a candidate set (e.g., a small subset of the full item set)~\cite{kusano2025revisiting, kim2024large}.}
Owing to LLM's strong reasoning and zero-/few-shot capabilities, LLM-based recommenders can handle cold-start items effectively and can also provide intuitive explanations that improve users’ understanding of the recommendation results.

To better adapt LLMs for the recommendation tasks, various \textit{retrieval-augmented generation} (RAG) techniques have been used~\cite{kim:hal-05561909, hu2026retrieval}.
Such methods typically focus on \textit{user retrieval}~\cite{wang2024whole, zhang2025adaptrec}, retrieving users similar to a target user and supplying their purchase histories together with the target user’s own history to the LLM.
% This process helps LLMs leverage recommendation-relevant knowledge without requiring expensive fine-tuning of LLMs.

However, user retrieval has several limitations.
First, retrieved users, although coarsely similar to the target user, may still include a considerable amount of irrelevant information, which can introduce noise.
Specifically, because individual users typically exhibit diverse interests, the retrieved purchase histories are likely to reflect the target user’s preferences only partially, with a large portion being less relevant. 
Second, this approach does not explicitly retrieve information with respect to candidate items, which are equally important to users in recommendations. 
Although the retrieved user histories may implicitly contain information relevant to candidate items, the lack of explicit conditioning can limit their effectiveness.

To address these limitations, we propose \textbf{\method}, whose key idea is to retrieve information relevant to individual items at a fine-grained level, instead of coarse-grained retrieval at the user level.
Especially, \method applies this retrieval not only to items in the user’s purchase history but also to each candidate item, %(which we refer to as query items), 
augmenting each item with descriptions of relevant items retrieved from the full item corpus.
This item-level augmentation (1) mitigates noise introduced by irrelevant information from partially relevant users in user-level augmentation, and (2) also offers explicit, recommendation-beneficial evidence for each candidate item.

As is intuitive, the success of \method depends on the retrieved items and their relevance.
However, a straightforward extension of a semantics-based retriever, a common design choice in natural language processing~\cite{karpukhin2020dense} that retrieves items with semantically similar descriptions, may be less effective for recommendation, as shown by its limited gains over the user-based approach ~\cite{wu2024coral} in Figure~\ref{fig:mainperformance}.

\begin{figure*}[t] 
    \centering
    \includegraphics[width=0.98\linewidth]{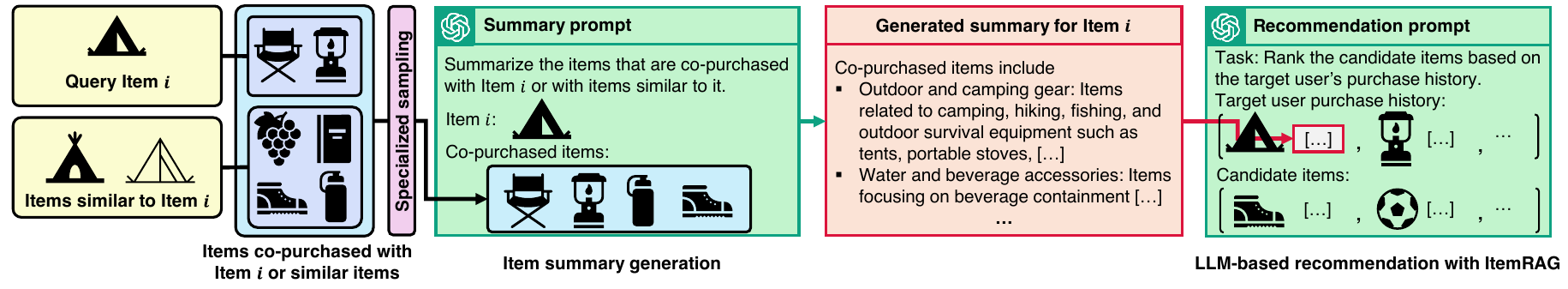} 
    \vspace{-1mm}
    \caption{\textit{An example case of \method, our item-based RAG method}.
    For retrieving relevant items for item $i$, we first identify items that are co-purchased with (1) item $i$ itself and/or (2) items whose textual descriptions are similar to that of item $i$. 
    Then, we sample a specified number of items from this pool, with selection probabilities proportional to their co-purchase frequencies with item $i$.
    Subsequently, we prompt an LLM to generate a summary of the sampled items and incorporate the summary into the final recommendation prompt, guiding the LLM to understand the co-purchase patterns among items.
    }
    \label{fig:mainfigure}
\end{figure*}

To enable recommendation-aligned retrieval, \method leverages item co-purchase relations alongside semantic information through a carefully designed retrieval strategy, which leads to substantial empirical performance gains (see Figure~\ref{fig:mainperformance}).
Specifically, to reduce the impact of less-relevant, incidentally co-purchased items, \method selectively samples retrieved items in proportion to their co-purchase frequencies, rather than using all retrieved items.
Moreover, to better handle cold-start items with few or no co-purchase neighbors, \method expands the retrieval set by additionally retrieving items that were co-purchased not with the query item itself, but with items that are semantically similar to it.

%\method samples a predefined number of 

%By retrieving items co-purchased with the query items, \method enables an LLM to understand the co-purchase patterns among items, which are often beneficial for the recommendations.

Through extensive experiments, we demonstrate the effectiveness of \method in both (1) standard LLM-based recommendation and (2) cold-start item recommendation. In particular, \method consistently outperforms user-based RAG baselines in both settings, yielding up to 11\% gains on the Toys dataset (in terms of Hit-Ratio@1) over the strongest user-based RAG baseline.

Our key contributions are summarized below:
\begin{itemize}[leftmargin=*]
    \item \textbf{New concept.} We propose item-based RAG for LLM-based recommendation, a fine-grained approach that enriches the descriptions of each target-user purchased item and candidate item.
    \item \textbf{New method.} We propose \method, an item-based RAG approach that integrates co-purchase relations and semantic information for relevant item retrieval.
    \item \textbf{Strong performance.} \method consistently outperforms user-based RAG baselines in both (1) standard LLM-based recommendation and (2) cold-start item recommendation.
\end{itemize}
Supplementary materials, code, and datasets are provided at \url{https://github.com/kswoo97/ItemRAG}.

    \section{Related work and preliminary}
    \label{sec:relatedwork}

In this section, we review related studies and present the preliminary concepts relevant to our work.

\noindent\textbf{Related work.} 
Thanks to their strong reasoning capabilities and ability to handle cold-start items, using LLMs as recommender systems has attracted substantial attention~\cite{kusano2025revisiting, kim2024large}.
To better adapt them to recommendation tasks, retrieval-augmented generation (RAG) techniques have been widely explored~\cite{hu2026retrieval, kim:hal-05561909}.
One line of work applies RAG in specific settings, including conversational recommendation~\cite{zhu2025collaborative, qiu2025graph} and knowledge–graph–based recommendation~\cite{wang2025knowledge, meng2025kerag_r}.
Another line of RAG work aims to improve general-purpose recommenders that rely primarily on user–item interactions~\cite{wang2024whole, zhang2025adaptrec, hu2026retrieval}, which is our focus.
% , which is a focus of our work.
They are mostly user-based approaches, with inherent limitations detailed in Section~\ref{sec:intro}.
% Note that, to our knowledge, \method is the first RAG approach that explicitly focuses on relevant item retrieval for LLM-based recommendation. 

% wang2024whole, zhang2025adaptrec, hu2026retrieval, wu2024coral

\noindent\textbf{Preliminary.} The user set and the item set are denoted by $\mathcal{U}$ and $\mathcal{I}$, respectively.
We consider a sequential recommendation setting, and therefore, each user $u \in \mathcal{U}$ is represented by her purchase history sequence: $\mathbf{u}\coloneqq[i^{(u)}_{1}, i^{(u)}_{2},\cdots ,i^{(u)}_{n_u}]$, where $i^{(u)}_{s}$ denotes the $s$-th item purchased by user $u$ and $n_u$ is the number of items purchased by $u$.
Each item $i \in \mathcal{I}$ has a text description $\mathscr{t}_i$, such as item title.

% \noindent\textbf{Preliminary.} The user set and the item set are denoted by $\mathcal{U}=\{u_{1},u_{2},\cdots u_{\vert \mathcal{U}\vert }\}$ and $\mathcal{I}=\{i_{1},i_{2},\cdots,i_{\vert \mathcal{I}\vert}\}$, respectively.
% We consider a sequential recommendation setting, and therefore, each user $u_{t} \in \mathcal{U}$ is represented by her purchase history sequence: $u_{t}\coloneqq[i_{(t,1)}, i_{(t,2)},\cdots ,i_{(t,n_{t})}]$.
% Each item $i_{k} \in \mathcal{I}$ is associated with a textual description $x^{(txt)}_{k}$, such as item title.

    \section{Proposed method}
    \label{sec:method}

In this section, we introduce \textbf{\method}, an item-based retrieval-augmented generation (RAG) method for LLM-based recommendation. 
We first give an overview of the \method pipeline (Section~\ref{subsec:overview}) and detail our retrieval strategy (Section~\ref{subsec:retrieval}).

%(\textbf{\underline{Item}}-based \textbf{\underline{R}}etrieval-\textbf{\underline{A}}ugmented \textbf{\underline{G}}eneration)

\subsection{Overall pipeline of \method}\label{subsec:overview}

We consider an LLM-based recommendation pipeline in which an LLM is given (1) the target user’s purchase history and (2) a set of candidate items. 
Then, the LLM is prompted to rank the candidates by their likelihood of being the target user’s next purchase.\footnote{Candidate items are often obtained by non-LLM-based recommender systems.}
Here, each item is represented by its textual description (e.g., item title).

By using \method, we \textit{enhance the description of each query item $i$} that is (1) purchased by the target user or (2) a recommendation candidate—by retrieving items relevant to item $i$.
Specifically, for each query item $i$, we retrieve items relevant to it, and then provide a summary of the retrieved items together with the original textual description of item $i$. % each query 
Notably, as discussed in Section~\ref{sec:intro}, this augments individual items in a fine-grained manner, in contrast to widely-used coarse target user-level augmentation.
In Section~\ref{subsec:retrieval}, we present key challenges in the relevant-item decision process and introduce our retrieval strategy that overcomes them.

\subsection{Retrieval strategy of \method}\label{subsec:retrieval}

{One} way to retrieve items related to a query item is to select those that are co-purchased with it.
However, this faces two challenges: (C1) it performs poorly—and may be infeasible—for cold-start query items with little or no co-purchase data, and (C2) some co-purchased items are incidental and thus weakly relevant.

To address the cold-start challenge (C1), for each query item $i$, we retrieve not only items co-purchased with $i$ but also items co-purchased with items whose text descriptions are similar to $i$.
Our rationale is that items similar to $i$ often share co-purchase patterns with $i$, giving strong complementary co-purchase {information} for $i$.

To address the weak-relevance challenge (C2), we score each query–retrieved item pair by its co-purchase frequency and use this score for the probability of the item being selected in the final retrieval.
Our rationale is that frequent co-purchases indicate strong relevance and {are less likely to be incidental.}

Based on these intuitions, we formally elaborate on our retrieval strategy.
We start with presenting two notations.
We denote a set of items purchased by user $u$ as $\mathcal{M}(u)$ (i.e., $\mathcal{M}(u) = \{i^{(u)}_{s}:s \in \{1,2,\cdots ,n_{u}\}\}$).
We also denote a set of items co-purchased with item $i$ as $\mathcal{N}(i)$ (i.e., $\mathcal{N}(i) = \{j : i\neq j, \exists u\in \mathcal{U} \ \text{s.t.} \ \{i,j\} \subseteq \mathcal{M}(u)\}$).

In retrieval for a query item $i$, we first find the top-$K$ whose textual descriptions are most similar to $i$; we denote this set as $\mathcal{T}(i)$.
Specifically, for each $j \in \mathcal{I}$, we encode its text description $\mathscr{t}_{j}$ via a pre-trained language model \texttt{LM}, obtaining the representation $\mathbf{z}_{j} \in \mathbb{R}^{d}$ (i.e., $\mathbf{z}_{j} = \texttt{LM}(\mathscr{t}_{j})$).
We then compute the cosine similarity between $i$ and each other item $j \in \mathcal{I} \setminus \{i\}$ (i.e., $(\mathbf{z}^{T}_{i}\mathbf{z}_{j})/(\lVert \mathbf{z}_{i}\rVert_{2} \lVert \mathbf{z}_{j}\rVert_{2})$), and select the top-$K$ by similarity; the resulting set is $\mathcal{T}(i)$.

We subsequently derive a retrieval pool $\mathcal{P}(i)$ comprising items co-purchased with (1) item $i$ itself ($\mathcal{N}(i)$) and/or (2) items having similar descriptions to $i$ ($\mathcal{T}(i)$). 
Formally, the pool is defined as:
\begin{equation}\label{eq:retrievalpool}
    \mathcal{P}(i) = \mathcal{N}(i) \cup  \{j : \exists q \in \mathcal{T}(i) \text{ s.t. } j \in \mathcal{N}(q)\}.
\end{equation}

% \begin{equation}\label{eq:retrievalpool}
%     \mathcal{P}(i_{k}) = \mathcal{N}(i_{k}) \cup \left(\bigcup_{i_{q} \in \mathcal{T}(i_{k})} \mathcal{N}(i_{q})\right).
% \end{equation}

% \begin{equation}\label{eq:retrievalpool}
%     \mathcal{P}(i_{k}) = \mathcal{N}(i_{k}) \cup \left(\bigcup_{i_{q} \in \mathcal{T}(i_{k})} \mathcal{N}(i_{q})\right).
% \end{equation}

After, instead of retrieving all the items within $\mathcal{P}(i)$, we sample $N$ number of items proportional to co-purchase frequencies.
Formally, let a co-purchase frequency of items $i$ and $j$ as $c_{ij}$ (i.e., $c_{ij} = \sum_{u \in \mathcal{U}} \mathbf{1}[\{i,j\} \in \mathcal{M}(u)]$, where $\mathbf{1}[\cdot]$ is an indicator function).
Then, a sampling weight $w_{ij}$ of item $j$ being retrieved for query item $i$ is defined as:
$w_{ij} = c_{ij} + \frac{1}{\vert \mathcal{T}(i)\vert} \sum_{q \in \mathcal{T}(i)}c_{qj},$
% \begin{equation}\label{eq:samplingweight}
%     w_{ij} = c_{ij} + \frac{1}{\vert \mathcal{T}(i)\vert} \sum_{q \in \mathcal{T}(i)}c_{qj},
% \end{equation}
where $c_{ij}$ denotes the co-purchase frequency between items $i$ and $j$, and the rest indicates the mean of co-purchase frequencies between item $j$ and items that are semantically similar to item $i$.

Subsequently, we sample $N$ items from the retrieval pool $\mathcal{P}(i)$ (Eq.~\eqref{eq:retrievalpool}), where each item $j \in \mathcal{P}(i)$ is drawn with the probability of $w_{ij}/(\sum_{q \in \mathcal{P}(i)}w_{iq})$. %(Eq.~\eqref{eq:samplingweight})
Lastly, we prompt an LLM to summarize the sampled items and append this summary to the original description of item $i$, helping the LLM-based recommender capture co-purchase information of item $i$.
Note that the co-purchase summary generation is independent of the target user; thus, the summary for a given item can be used for different target users.
%\vspace{-2mm}

\begin{table*}[t]
\centering
\small
\setlength{\tabcolsep}{3pt} % default ~6pt
\caption{\textit{(RQ1\&4) LLM-based recommendation performance}.
All metrics are multiplied by 100 for better readability. 
H@K and N@K denote Hit-Ratio@K and NDCG@K, respectively. 
We do not report N@1, since it is equal to H@1.
Best results are highlighted with a \gframe{\textbf{green}} box, and * indicates that \method achieves statistically significant improvement over the corresponding baseline at the 0.05 significance level.
Notably, \method outperforms the baseline methods in 18 out of 20 cases.}
\vspace{-1mm}
\label{tab:performance}
{
\renewcommand{\arraystretch}{1.0}
\resizebox{1.0\linewidth}{!}{
\begin{tabular}{l| ccccc | ccccc | ccccc | ccccc}
\toprule

\multirow{2}{*}{Methods} & \multicolumn{5}{c|}{Beauty \& Personal Care} & \multicolumn{5}{c|}{Toys \& Games} & \multicolumn{5}{c|}{Sports \& Outdoors} & \multicolumn{5}{c}{Arts, Crafts \& Sewing} \\

& H@1 & H@3 & H@5 & N@3 & N@5 & H@1 & H@3 & H@5 & N@3 & N@5 & H@1 & H@3 & H@5 & N@3 & N@5 & H@1 & H@3 & H@5 & N@3 & N@5\\

\midrule
\midrule

LightGCN~\cite{he2020lightgcn} 
& 35.8* & 62.8* & 78.8* & 51.2* & 57.9* 
& 40.9* & 67.7* & 81.5* & 56.4* & 62.1*
& 45.3* & 70.6* & 83.3* & 60.2* & 65.5* 
& 51.7* & 78.2* & 88.0* & 67.1* & 71.1* \\

LightGCN++~\cite{lee2024revisiting} 
& 38.3* & 65.2* & 80.2* & 53.4* & 59.7* 
& 42.2* & 68.5* & 82.5* & 57.3* & 63.1*  
& 46.0* & 70.8* & 84.0* & 60.5* & 65.9* 
& 53.0* & 79.9* & 88.8* & 68.2* & 72.2* \\

SASRec~\cite{kang2018self} 
& 35.2* & 62.6* & 78.7* & 51.9* & 57.5* 
& 32.4* & 58.5* & 75.1* & 47.0 & 54.6* 
& 44.0* & 68.9* & 82.1* & 57.9* & 63.3* 
& 47.4* & 72.9* & 86.5* & 62.5* & 69.6* \\

BERT4Rec~\cite{sun2019bert4rec} 
& 36.1* & 63.2* & 78.9* & 52.3* & 57.7*  
& 32.0* & 58.7* & 75.8* & 47.1* & 55.5*
& 44.2* & 70.5* & 83.8* & 59.4* & 65.3* 
& 48.2* & 74.6* & 87.1* & 63.3* & 69.9* \\

\midrule

Zero-shot 
& 34.6* & 57.7* & 72.9* & 48.0* & 54.2* 
& 39.3* & 62.5* & 76.5* & 52.7* & 58.5* 
& 43.4* & 66.6* & 81.6* & 56.7* & 62.9* 
& 45.7* & 72.2* & 83.8* & 61.1* & 65.6* \\

\midrule

ICL~\cite{wang2024whole} 
& 37.7* & 61.9* & 76.7* & 51.5* & 57.5* 
& 41.3* & 66.4* & 80.6* & 55.7* & 61.1* 
& 46.4* & 70.2* & 84.1* & 59.8* & 65.7*  
& 46.3* & 75.3* & 86.7* & 63.1* & 67.8* \\

AdaptRec~\cite{zhang2025adaptrec} 
& 34.7* & 58.2* & 74.9* & 47.9* & 55.1*
& 37.8* & 63.3* & 77.5* & 52.2* & 58.1* 
& 44.4* & 67.8* & 83.2* & 57.5* & 63.6*  
& 46.4* & 74.4* & 85.8* & 62.9* & 67.9* \\

ReACT~\cite{hu2026retrieval} 
& 34.4* & 57.7* & 73.5* & 47.6* & 54.2* 
& 38.5* & 61.7* & 75.9* & 51.7* & 57.8* 
& 44.7* & 68.7* & 83.5* & 58.2* & 64.4* 
& 46.9* & 73.0* & 83.8* & 61.9* & 66.4* \\

CoRAL~\cite{wu2024coral} 
& 46.5* & 69.9* & 82.0* & 60.2* & 65.0* 
& 44.6* & 70.1* & 81.2* & 59.4* & 64.0* 
& 48.5* & 73.2* & 87.1* & 62.5* & 68.3*  
& 52.4* & 80.7* & 91.1* & 68.9* & 73.3* \\

\midrule

w/o cand-aug.
& 36.9* & 59.2* & 73.8* & 49.8* & 55.8*
& 42.1* & 63.7* & 78.9* & 54.6* & 60.8*
& 44.8* & 68.2* & 81.8* & 57.5* & 63.3*
& 47.3* & 72.8* & 84.4* & 62.0* & 66.8* \\

w/o co-purch.
& 46.1* & 68.9* & 82.1* & 59.1* & 64.7* 
& 46.4* & 70.8* & 81.9* & 60.6* & 65.1* 
& 50.6* & 72.9* & 86.2* & 63.6* & 68.9* 
& 54.6* & 80.1* & 89.7* & 69.5* & 73.3*  \\

w/o sim-items 
& 47.5* & 70.2* & 82.8* & 60.8* & 66.3* 
& \best 49.8 & 71.6* & 83.9* & 62.4* & 67.7* 
& 50.2* & 74.0* & 87.7* & 64.5 & 70.0  
& 56.0* & 82.1* & 91.1* & 71.0* & 74.6* \\

w/o co-freq. 
& 47.7* & 70.1 & \best 83.9 & 60.7* & 66.6
& 48.1* & 73.1* & 83.5* & 63.0* & 67.0*  % Toy
& 50.7* & 74.1 & 87.9* & 64.0 & 69.8*  % Sport
& 56.2* & 82.2* & 91.0* & 71.3* & 75.0* \\ % Art

\midrule

\method 
& \best 49.1 & \best 71.0 &       83.8 & \best 61.5 & \best 66.9
&       49.5 & \best 74.2 & \best 85.0 & \best 63.8 & \best 68.1
& \best 51.6 & \best 74.7 & \best 88.3 & \best 64.6 & \best 70.3
& \best 57.8 & \best 82.6 & \best 91.9 & \best 72.3 & \best 76.1
\\

% Improvement & +00.0\% & +00.0\% & +00.0\% & +00.0\% & +00.0\% & +00.0\% & +00.0\%  & +00.0\% & +00.0\% & +00.0\% & +00.0\% & +00.0\% & +00.0\% & +00.0\% & +00.0\% & +00.0\% 
% & +00.0\% & +00.0\% & +00.0\% & +00.0\%\\

\bottomrule
\end{tabular}
}
}
\end{table*}

    \section{Experiment}
    \label{sec:experiment}

In this section, we analyze the effectiveness of \method in the LLM-based recommendation tasks.
We answer the questions below:
\begin{enumerate}[label={RQ\arabic*.}, leftmargin=*]
  \item How effective is \method for LLM-based recommendation?
  \item How accurate is \method at recommending cold-start items?
  \item Do LLM-based recommender systems make effective use of the item information retrieved by \method?
  % When the item information retrieved by \method is given, do LLM-based recommender systems use it?
  \item Do all \method key components contribute to performance?
\end{enumerate}

%\vspace{-2mm}

\subsection{Experimental setting}\label{subsec:protocol}

\noindent\textbf{Datasets and evaluation protocol.}
We use four domains from the latest Amazon Reviews dataset~\cite{hou2024bridging}: Sports \& Outdoors (Sports), Toys \& Games (Toys), Beauty \& Personal Care (Beauty), and Arts, Crafts \& Sewing (Arts).
Further details, including preprocessing steps and dataset statistics, are provided in Appendix~\cite{anongithub}.
For evaluation, following prior work~\cite{wu2024coral, kusano2025revisiting}, we use a leave-one-out protocol: for each user, the last purchased item is held out for testing, and the remaining history is used as input.
Also, following~\cite{kusano2025revisiting}, we prompt the LLM to rank 10 candidate items for the target user’s next purchase; the set includes 1 ground-truth item and 9 randomly sampled items, and results with larger candidate sets are reported in Appendix~\cite{anongithub}.
We run each experiment three times and report the mean metrics. 
We further conduct a Wilcoxon signed-rank test between \method and each baseline to assess statistical significance.
In addition, we provide an inference runtime analysis of \method in the appendix~\cite{anongithub}.

\noindent\textbf{Baseline methods and \method.}
For comparison, we use 9 baseline methods: two graph-based models (LightGCN~\cite{he2020lightgcn} and LightGCN++~\cite{lee2024revisiting}), two sequential models (SASRec~\cite{kang2018self} and BERT4Rec~\cite{sun2019bert4rec}), one naive zero-shot LLM-based recommender, and four user-based RAG methods (ICL~\cite{wang2024whole}, AdaptRec~\cite{zhang2025adaptrec}, ReACT~\cite{hu2026retrieval}, and CoRAL~\cite{wu2024coral}).
For LLM-based methods, we use GPT-4.1-mini as the backbone.
For the retrieval process in \method, we use $5$ similar items per item and sample $50$ items in the final retrieval set. 
Further details on baselines, hyperparameters, and prompts are given in Appendix~\cite{anongithub}.

\vspace{-1mm}

\subsection{RQ1. Standard LLM-based recommendation}\label{subsec:standardsetting}

\noindent\textbf{Setup.}
For each method, we construct the training data and retrieval database from users’ purchase histories after withholding each user’s last interaction, which is reserved exclusively for evaluation.
For testing, we evaluate each method on $1{,}000$ randomly sampled users under the evaluation protocol detailed in Section~\ref{subsec:protocol}.

\noindent\textbf{Result.}
As shown in Table~\ref{tab:performance}, \method outperforms all the baseline methods in 18 out of 20 settings.
Two points stand out.
First, \method consistently improves the naive zero-shot LLM recommender, by up to 42\% in Hit-Ratio@1 on the Beauty \& Personal Care dataset.
Second, \method outperforms user-based RAG methods, outperforming the strongest baseline (CoRAL) by up to 11\% in terms of Hit-Ratio@1 on the Toys \& Games dataset.

%\vspace{-1mm}

\subsection{RQ2. Cold-start item recommendation}

\noindent\textbf{Setup.} Given that the strength in cold-start item recommendation is the primary promise of LLM-based recommenders~\cite{wu2024survey}, we evaluate each method under an item cold-start setup. 
Specifically, for the $1,000$ sampled users described in Section~\ref{subsec:standardsetting}, we remove the ground-truth test items for the users—together with all interactions involving that item—from both the training set and the retrieval database, making the corresponding items cold-start.\footnote{Since the original dataset contains very few items that first appear in the test set, we use the modified dataset instead.} % —fewer than 30 across all datasets—
We then evaluate each method's ability to recommend such items to the corresponding users.
Since the learning-based baselines we use cannot handle unseen (cold-start) items, we focus on LLM-based approaches.
Other settings remain the same as in Section~\ref{subsec:standardsetting}.

\begin{table*}[t]
\centering
\small
\setlength{\tabcolsep}{4.0pt} % default ~6pt
\caption{\textit{(RQ2) Cold-start item recommendation performance}.
All metrics are multiplied by 100 for better readability. 
H@K and N@K denote Hit-Ratio@K and NDCG@K, respectively. 
Best results are highlighted with a \gframe{\textbf{green}} box, and * indicates that \method achieves statistically significant improvements over the corresponding baseline at the 0.05 significance level. 
Notably, \method outperforms the baseline methods in every case.}
\label{tab:coldperformance}
{
\renewcommand{\arraystretch}{1.0}
\resizebox{\linewidth}{!}
{
\begin{tabular}{l| ccccc | ccccc | ccccc | ccccc}
\toprule

\multirow{2}{*}{Methods} & \multicolumn{5}{c|}{Beauty \& Personal care} & \multicolumn{5}{c|}{Toys \& Games} & \multicolumn{5}{c|}{Sports \& Outdoors} & \multicolumn{5}{c}{Arts, Crafts \& Sewing} \\
& H@1 & H@3 & H@5 & N@3 & N@5 
& H@1 & H@3 & H@5 & N@3 & N@5
& H@1 & H@3 & H@5 & N@3 & N@5 
& H@1 & H@3 & H@5 & N@3 & N@5 \\

\midrule
\midrule

Zero-shot 
& 34.6* & 57.7* & 72.9* & 48.0* & 54.2* 
& 39.3* & 62.5* & 76.5* & 52.7* & 58.5* 
& 43.4* & 66.6* & 81.6* & 56.7* & 62.9* 
& 45.7* & 72.2* & 83.8* & 61.1* & 65.6* \\

\midrule

ICL~\cite{wang2024whole}  
& 37.1* & 61.6* & 76.4* & 51.3* & 56.9* 
& 40.5* & 66.1* & 79.1* & 55.1* & 60.7*  
& 45.6* & 69.7* & 83.9* & 59.4* & 65.3* 
& 45.0* & 74.3* & 85.8* & 61.9* & 66.6* \\

AdaptRec~\cite{zhang2025adaptrec}
& 37.0* & 60.4* & 75.5* & 50.1* & 56.1* 
& 38.2* & 64.4* & 78.6* & 53.3* & 59.7*  
& 45.2* & 68.9* & 83.0* & 58.8* & 64.6* 
& 49.5* & 73.5* & 85.0* & 62.4* & 67.1* \\

ReACT~\cite{hu2026retrieval}
& 35.3* & 58.4* & 72.0* & 50.4* & 53.8* 
& 39.0* & 61.9* & 76.2* & 52.2* & 57.9* 
& 42.9* & 66.4* & 81.1* & 56.0* & 62.2* 
& 47.2* & 72.7* & 84.4* & 62.0* & 66.8* \\

CoRAL~\cite{wu2024coral}
& 38.5* & 60.9* & 73.3* & 51.5* & 56.8* 
& 39.0* & 61.2* & 74.0* & 51.7* & 57.0*  
& 45.0* & 68.2* & 81.4* & 58.6* & 63.4* 
& 46.4* & 73.4* & 83.9* & 61.8* & 66.3* \\

\midrule

\method
& \best 47.7 & \best 70.7 & \best 83.7 & \best 60.9 & \best 66.2 
& \best 48.7 & \best 71.9 & \best 83.2 & \best 61.9 & \best 66.7  
& \best 51.3 & \best 74.7 & \best 88.0 & \best 64.3 & \best 70.2 
& \best 57.9 & \best 82.7 & \best 91.4 & \best 72.3 & \best 75.9 \\

% Improvement & +00.0\% & +00.0\% & +00.0\% & +00.0\% & +00.0\% & +00.0\% & +00.0\% & +00.0\% & +00.0\% & +00.0\% & +00.0\% & +00.0\% & +00.0\% & +00.0\% & +00.0\% & +00.0\% \\

\bottomrule
\end{tabular}
}
}
\vspace{2mm}
\end{table*}

\begin{figure}[t] 
  \centering
  \includegraphics[width=1.0\linewidth]{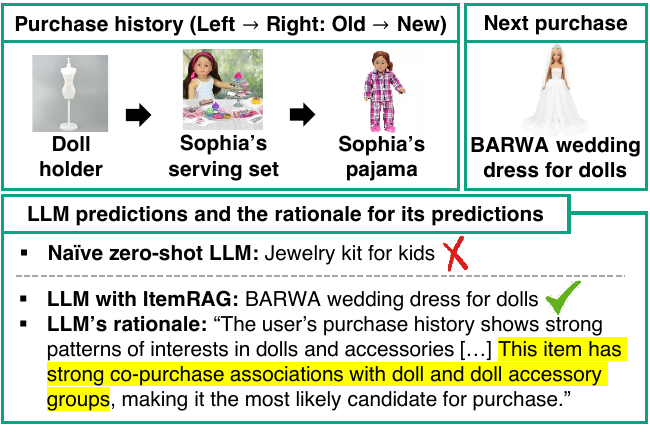} 
  \caption{\textit{(RQ3) Case study.}
  While the naive zero-shot LLM-based recommender fails, augmenting it with co-purchase information retrieved by \method—information the model explicitly uses—yields an accurate recommendation.}
  \label{fig:casestudy}
\end{figure}

% \noindent\textbf{Result.}
% As shown in Table~\ref{tab:coldperformance}, \method outperforms the baseline methods in all the cases, demonstrating its strong performance in recommending cold-start items.
% \blue{Notably, \method's performance drop compared to the standard setting (Table~\ref{tab:performance}) is only 1.3\% on average across all the cases, supporting its strong performance in cold-start.}

\noindent\textbf{Result.}
As shown in Table~\ref{tab:coldperformance}, \method outperforms the baseline methods in all the cases, demonstrating its strong performance in recommending cold-start items. 
Notably, its performance decreases by only 1\% on average relative to the standard setting (Table~\ref{tab:performance}), suggesting that it remains effective in cold-start scenarios.

\subsection{RQ3. Case study}

\noindent\textbf{Setup.}
We examine whether the LLM-based recommender system leverages the item information retrieved by \method. 
To this end, on the Toys \& Games dataset, we run a case study in which the LLM is prompted to give the rationale behind its recommendations. 
Additional cases in other datasets are reported in Appendix~\cite{anongithub}.

\noindent\textbf{Result.}
Figure~\ref{fig:casestudy} presents a case where a naive zero-shot LLM-based recommender fails to provide an accurate recommendation.
When the prompt is augmented with co-purchase information retrieved by ItemRAG, the LLM (1) recommends the correct item and (2) explicitly notes in its rationale that it relied on the retrieved co-purchase signals.
This result suggests that \method's retrieved information is indeed used and beneficial for improving performance.

\subsection{RQ4. Ablation study}

\noindent\textbf{Setup.}
We assess the necessity of \method’s key components by using four variants below: 
\begin{enumerate}[label=(V\arabic*), leftmargin=*]
  \item w/o cand-aug: Does not augment candidate item descriptions.
  \item w/o co-purch: Retrieves textually similar items instead of using co-purchase relations.
  \item w/o sim-items: Uses its own co-purchase information for each item during retrieval
  \item w/o co-freq: Replaces frequency-weighted sampling with uniform sampling.
\end{enumerate}
% (i) a variant that only uses its own co-purchase information for each item during retrieval ({w/o sim-items}), and 
% (ii) a variant that replaces frequency-based sampling weights with uniform sampling ({w/o co-purch.}).

\noindent\textbf{Result.}
As shown in Table~\ref{tab:performance}, the four variants underperform \method in 18 out of 20 settings, demonstrating the effectiveness of the \method's key components in LLM-based recommendation.

    \section{Conclusion and discussion}
    \label{sec:conclusion}
    In this work, we introduce \method, an item-based RAG technique for LLM-based recommendation.
{The key idea is to augment individual items in the target user’s purchase history or the candidate set, instead of relying on coarse user-level augmentation. 
Especially, its carefully designed retrieval strategy, guided by co-purchase information, retrieves items that are recommendation-relevant rather than merely semantically similar, and also augments cold-start items.}
Through extensive experiments, we demonstrate the effectiveness of \method for LLM-based recommendation and cold-start item recommendation.
One limitation of \method is that incorporating additional retrieved information can increase the length of the input prompt to the LLM recommender, which in turn leads to higher API costs and longer inference time.
Reducing the token usage by the retrieved information is thus an important direction for future work, especially for practical deployment.

% Supplementary materials, code, and datasets are provided at \url{https://github.com/kswoo97/ItemRAG}.

    \vspace{1mm}
    \noindent\textbf{Acknowledgements.} This work was partly supported by the National Research Foundation of Korea (NRF) grant funded by the Korea government (MSIT) (No. RS-2024-00406985, 40\%).
    This work was partly supported by Institute of Information \& Communications Technology Planning \& Evaluation (IITP) grant funded by the Korea government (MSIT) (No. RS-2022-II220871, Development of AI Autonomy and Knowledge Enhancement for AI Agent Collaboration, 50\%)
    (No. RS-2019-II190075, Artificial Intelligence Graduate School Program (KAIST), 10\%).

    %\newpage
    
    \bibliographystyle{ACM-Reference-Format}
    \balance
	\bibliography{000Ref}
    
    % \newpage
    
    % \appendix

    % {\LARGE\noindent\textbf{Appendix}}

    % \section{Details of parameter-efficient fine-tuning}
    % \label{app:peftdetail}
    % \input{901PEFTDetail}

    % \section{Details of dataset}
    % \label{app:datadetail}
    % \input{902DatasetDetail}

    % \section{Details of experiment}
    % \label{app:experimentdetail}
    % \input{903ExperimentDetail}

    % \section{Details of \method}
    % \label{app:method}
    % \input{904MethodDetail}

    % \section{Audience Interactivity}
    % \label{sec:interacitivity}
    %     \input{906Interactivity}

    % \section{societal Impact}
    % \label{sec:impact}
    %     \input{907Impact}

% \smallsection{Acknowledgements}
% This work was partly supported by the National Research Foundation of Korea (NRF) grant funded
% by the Korea government (MSIT) (No. RS-2024-00406985).
% This work was partly supported by Institute of Information \& Communications Technology Planning \& Evaluation (IITP) grant funded by the Korea government (MSIT) (No. RS-2024-00438638, EntireDB2AI: Foundations and Software for Comprehensive Deep Representation Learning and Prediction on Entire Relational Databases) (No. RS-2019-II190075, Artificial Intelligence Graduate School Program (KAIST)). This work has been partially supported by the spoke ``FutureHPC \& BigData” of the ICSC – Centro Nazionale di Ricerca in High-Performance Computing, Big Data and Quantum Computing funded by European Union – NextGenerationEU.
        %\newpage

%        \clearpage %do not use this here
        % \newpage 
\end{document}